\documentclass{ws-ijmpc}

\begin{document}

\markboth{Nuno Crokidakis}
{Emergence of moderate opinions as a consequence of group pressure}

\catchline{}{}{}{}{}



\title{Emergence of moderate opinions as a consequence of group pressure}

\author{Nuno Crokidakis}

\address{Instituto de F\'isica, Universidade Federal Fluminense \\
Niter\'oi/RJ, Brazil\\
nuno@mail.if.uff.br}

\maketitle

\begin{history}
\received{Day Month Year}
\revised{Day Month Year}
\end{history}

\begin{abstract}
In this work we study a continuous opinion dynamics model considering 3-agent interactions and group pressure. Agents interact in a fully-connected population, and two parameters govern the dynamics: the agents' convictions $\lambda$, that are homogeneous in the population, and the group pressure $p$. Stochastic parameters also drive the interactions. Our analytical and numerical results indicate that the model undergoes symmetry-breaking transitions at distinct critical points $\lambda_{c}$ for any value of $p<p^{*}=2/3$, i.e., the transition can be suppressed for sufficiently high group pressure. Such transition separates two phases: for any $\lambda \leq \lambda_{c}$, the order parameter $O$ is identically null ($O=0$, a symmetric, absorbing phase), while for $\lambda>\lambda_{c}$, we have $O>0$, i.e., a symmetry-broken phase (ferromagnetic). The numerical simulations also reveal that the increase of group pressure leads to a wider distribution of opinions, decreasing the extremism in the population.


\end{abstract}

\ccode{PACS Nos.: 05.10.-a, 05.70.Jk, 87.23.Ge, 89.75.Fb}

\section{Introduction}

\qquad The study of dynamics of social interactions is a topic of intense research in Statistical Physics of Complex Systems. Indeed, there is a considerable number of papers on this subject that has been published in the last years (see \cite{galam_book,sen_book,pmco_book,rmp}). One of the reasons for this interest is that even simple models can exhibit a complex collective behavior that emerges from the interactions among individuals in a given network of contacts. Usually, those models exhibit phase transitions and rich critical phenomena, which justifies the theoretical interest of physicists in the study of social dynamics.

Opinion dynamics is subject of great interest for the physics community. In recent years a series of papers considers different microscopic rules of interaction, inflow and outflow dynamics, distinct networks of contacts, continuous and discrete opinions, presence of activists and many others \cite{gastner,han,wang,gracia,abramiuk,weber,mori,abid,trump,javarone_galam,nuno_allan,celia_allan,nuno_pmco,bjp,sznajd_app,nuno_jorge,cheng_19,javarone_galam2,Bottcher_PLOS}. In addition, a general unifying framework was developed to study social contagions in the context of complex networks \cite{Bottcher_PRL}. The interest in such systems range from theoretical questions like phase transitions, universality and effects of disorder, to practical concerns like comparison with real data (elections or referendums, for example) and predictions of specific behaviors.

Recently, it was proposed a model based on bounded confidence \cite{deffuant} that considered the influence of group pressure on opinion formation \cite{cheng_19}. The authors find a rich beahvior, and the results suggest that a group with all individuals facing group pressure can reach a consensus in finite time, due to the competition between bounded confidence and group pressure. In this work we propose a model based on kinetic exchange opinion dynamics, also considering the mentioned mechanism of group pressure. The analytical and numerical results suggest that the model undergoes nonequilibrium phase transitions for certain range of the model's parameters, and that such transition is suppressed for sufficiently high group pressure. In addition, the increase of the group pressure decreases the extremism in the population, even if the agents' convictions are high.

This work is organized as follows. In Section 2 we present the model and define the microscopic rules that govern the dynamics. The analytical and numerical results are discussed in Section 3. Our conclusions are presented in Section 4.


\section{Model}

\qquad We consider a fully-connected population wit $N$ agents or individuals. Each individual $i$ carries an opinion $o_{i}$, given by a continuous variable in the real range $[-1,1]$, in order to represent the possible shades of individual’s attitudes against or for the topic under discussion \cite{deffuant,nuno_celia_pre17,marlon,fan,hk,deffuant3,lorenz,coda,jstat,lccc,biswas_conf,brugna,balenzuela}. Opinions tending to $o=\pm 1$ indicate extremist individuals, while opinions $o\approx 0$ mean neutral or undecided ones. 

Next, we present our kinetic exchange opinion model with group pressure that describes the evolution of an agent's expressed opinion while under pressure to conform with the public opinions in the group. The model is based in the so-called LCCC (Lallouache-Chakrabarti-Chakraborti-Chakrabarti) model \cite{lccc}, and in a recent proposed model considering group pressure \cite{cheng_19}. Initially, the opinion of each agent is randomly chosen from a uniform distribution $[-1,1]$. At each interaction, we choose at random 3 individuals, say $i$, $j$ and $k$, to form a group. The opinion of a given agent $i$ in the next time step $t$ evolves as
\begin{equation}\label{eq1}
o_{i}(t+1) = (1-p)\,[\lambda\,o_{i}(t) + \lambda\,\epsilon_{t}\,o_{j}(t) + \lambda\,\epsilon_{t'}\,o_{k}(t)] + p\,\epsilon_{t''}\,o_{avg}(t)  ~.
\end{equation}
\noindent
In Eq. \eqref{eq1} , $\epsilon_{t}$, $\epsilon_{t'}$ and $\epsilon_{t''}$ are stochastic random variables, changing with time (annealed variables). These variables are uniformly distributed in the range $[0,1]$, as in Ref. \cite{lccc}. The parameter $\lambda$ represents the conviction of each agent, and we assumed for simplicity that all agentes are homogeneous, i.e., they have the same conviction $\lambda$. This parameter is defined in the range $0\leq \lambda\leq 1$, as in the original formulation of the LCCC model \cite{lccc}. We denote the group pressure as $p$, while the resilience to this pressure is represented as $1-p$. In addition, we have definied
\begin{equation}\label{eq2}
o_{avg}(t) = \frac{o_{i}(t)+o_{j}(t)+o_{k}(t)}{3}
\end{equation}
\noindent
as the average opinion of the group formed by the 3 agents $(i,j,k)$. Notice that Eq. (\ref{eq1}) changes only the opinion of agent $i$ at step $t+1$. $N$ of such updates define the unit of time. The opinions are bounded, i.e., $-1 \leq o_{i}(t) \leq 1$. As one can see in Eq. \eqref{eq1}, the case $p=0$ (no group pressure) corresponds to a pure 3-group interaction, that was not studied before in the context of the LCCC model \cite{biswas_conf}.

To characterize the coherence of the collective state of the population, we employ  
\begin{equation} \label{eq3}
O  =   \frac{1}{N}\left|\sum_{i=1}^{N} o_{i}\right|  ~. 
\end{equation}
Notice that this is a kind of order parameter that plays the role of  the ``magnetization per spin'' in  magnetic systems. It is sensitive to the unbalance between positive and negative opinions. A collective state with a significantly non-null value of $O$ means a symmetry-broken distribution of opinions.  Therefore,  the debate has a clear result, be extremist or moderate. In the next section we discuss our analytical and numerical results.


\section{Results}

\qquad The LCCC model and its extensions present continuous symmetry-breaking nonequilibrium phase transitions \cite{biswas_conf}. In such case, numerical simulation reveals that the system goes into either of two possible phases: for any $\lambda \leq \lambda_{c}$, $O=0, \forall i$ (a symmetric phase), while for $\lambda>\lambda_{c}$, $O>0$,  a symmetry-broken phase, with critical points $\lambda_{c}$ that depend on the formulation of the specific extension of the LCCC model (for a brief review, see \cite{biswas_conf}). Usually one of the opinion sides, positive or negative, disappears of the population in the long-time limit in the symmetry-broken (ferromagnetic) phase.

Before we discuss the numerical results, a mean field calculation can be proposed for the fixed point $o^{*}$, following \cite{biswas_conf}. Considering such fixed point in Eqs. \eqref{eq1} and \eqref{eq2}, one obtains
\begin{equation}\label{eq4}
o^{*} = (1-p)\,[\lambda\,o^{*} + \lambda\,\langle\epsilon_{t}\rangle\,o^{*} + \lambda\,\langle\epsilon_{t'}\rangle\,o^{*}] + p\,\langle\epsilon_{t''}\rangle\,(o^{*}+o^{*}+o^{*})/3  ~.
\end{equation}
For uniform random distribution of $\epsilon$ in the range $[0,1]$, $\langle\epsilon\rangle=1/2$, thus we found
\begin{equation}\label{eq5}
\lambda_{c}(p) = \frac{2-p}{4(1-p)}  ~.
\end{equation}
\noindent
This result foresee that the model undergoes phase transitions for distinct values of the group pressure $p$. However, there is a range of values of $p$ for which the mentioned transition occurs. Since $\lambda_{c}$ grows for increasing values of $p$, and the conviction $\lambda$ is a parameter defined in the range $0\leq \lambda\leq 1$, we have a limiting value $p^{*}$ that is assoiated with the maximum value of $\lambda_{c}$, namely $\lambda_{c}=1$. Taking $\lambda_{c}(p^{*})=1$ in Eq. \eqref{eq5}, one obtains
\begin{equation}\label{eq6}
p^{*} = \frac{2}{3} \approx 0.67  ~.
\end{equation}
\noindent
In such case, if more than $67\%$ of the interactions take into account the overall group opinion, there is no ordering in the system for any value of the convition $\lambda$.

\begin{figure}[t]
\begin{center}
\vspace{6mm}
\includegraphics[width=0.6\textwidth,angle=0]{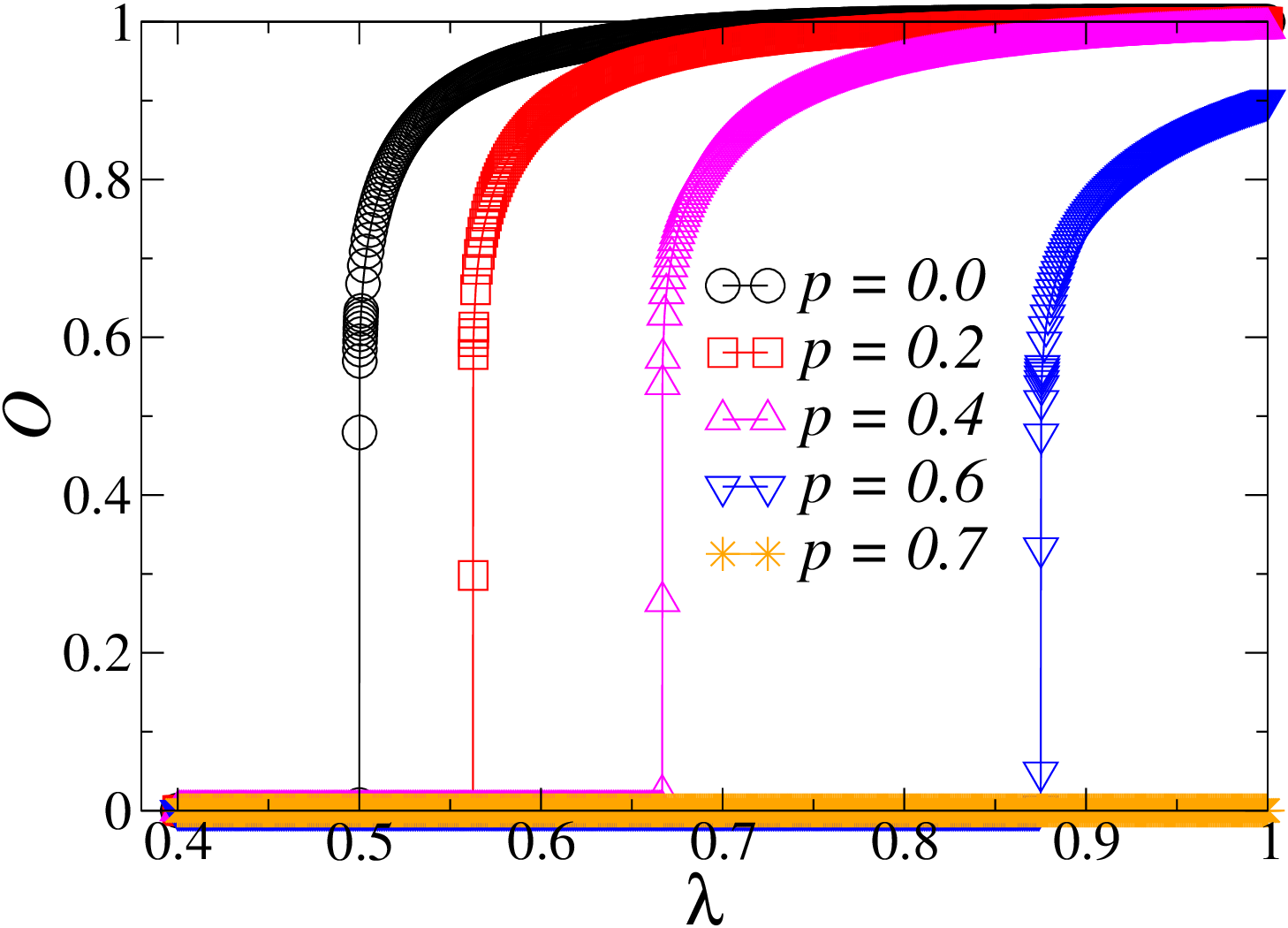}
\end{center}
\caption{(Color online) Order parameter $O$ as a function of $\lambda$ for typical values of the group pressure $p$. The system undergoes symmetry-breaking phase transitions at distinct critical points $\lambda_{c}(p)$, as discussed in the text. The population size is $N=10^{4}$, and data are accumulated over $100$ independent simulations}.
\label{fig1}
\end{figure}

To test those analytical predictions, we performed computer simulations of the model for populations of size $N=10000$. Fig. \ref{fig1} shows the stationay order parameter $O$, defined in Eq. \eqref{eq3}, as a function of $\lambda$ for typical values of the group pressure $p$. We can see that the model undergoes phase transition for some values of $p$, but for sufficiently large values like $p=0.7$ there is no ordering in the system, in agreement with the analytical result of Eq. \eqref{eq6}. This is a first consequence of the group pressure in the LCCC model with groups formed by three agents, and the following results clarify the effects of group pressure. For other comparison of analytical and numerical results, in the case of a pure 3-group interaction ($p=0$), the numerical results suggest that $\lambda_{c}\approx 0.5$, in agreement with the analytical result $\lambda_{c}(p=0)=1/2$ of Eq. \eqref{eq5}. 

\begin{figure}[t]
\begin{center}
\vspace{6mm}
\includegraphics[width=0.48\textwidth,angle=0]{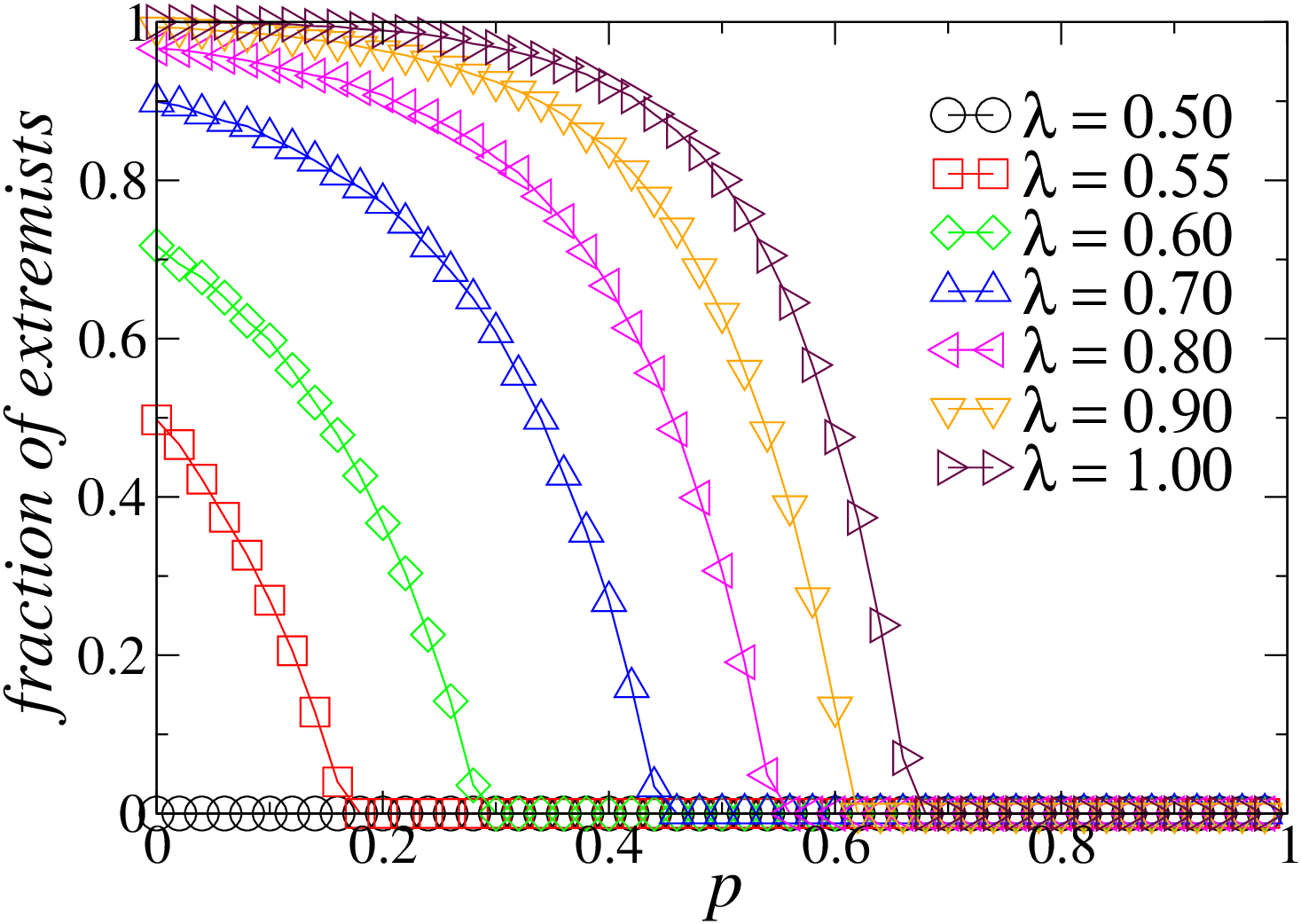}
\hspace{0.3cm}
\includegraphics[width=0.48\textwidth,angle=0]{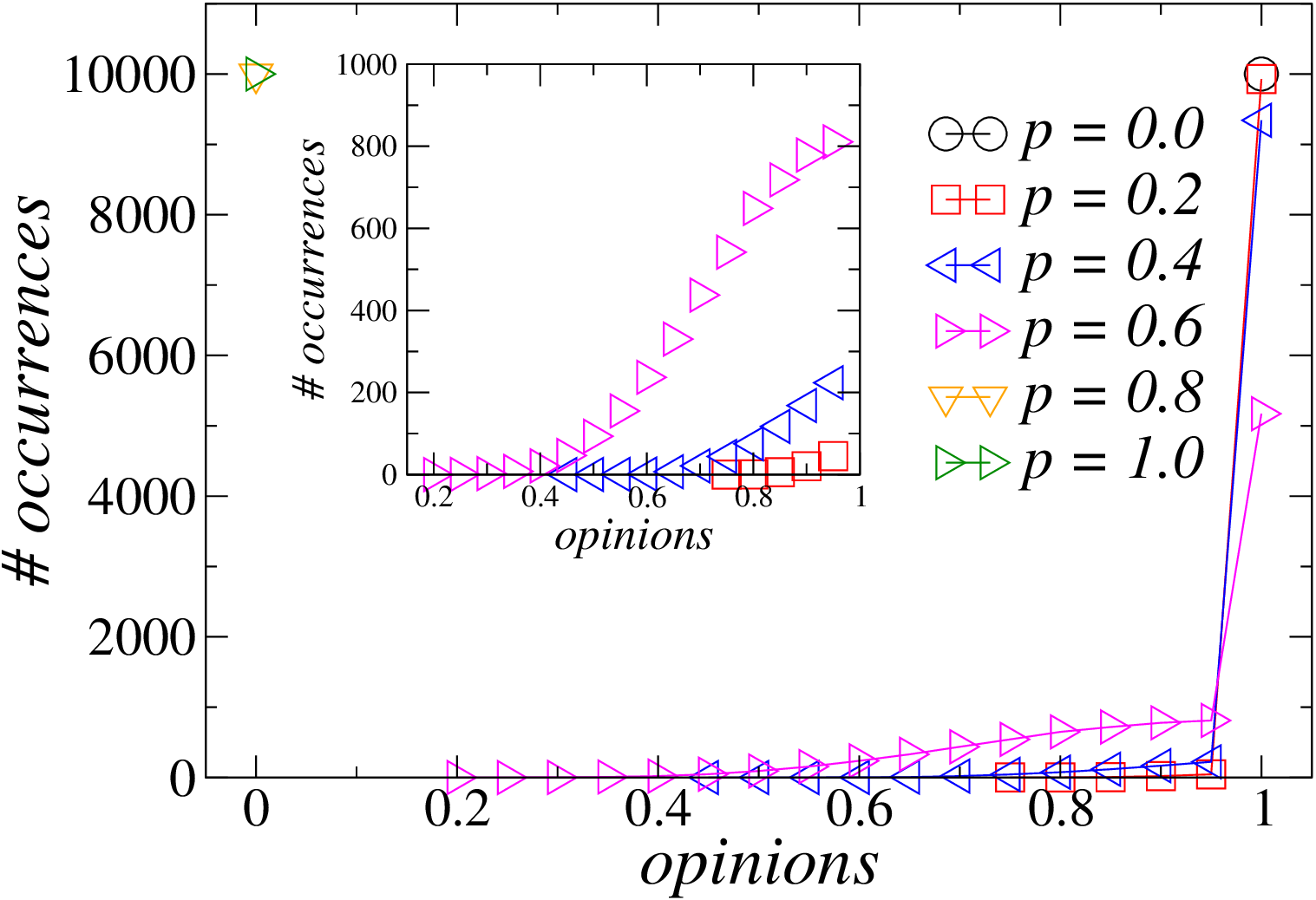}
\end{center}
\caption{(Color online) \textit{Left side}: Stationary fraction of agents with extreme opinions ($o=\pm 1$) as a function of the group pressure $p$, for typical values of $\lambda$. \textit{Right side}: Histograms of opinions, in the stationary states, for $\lambda=1.0$ and typical values of $p$. In the inset we show a zoom of the main frame, excluding the extremist agents with majority opinions. The population size is $N=10^{4}$, and data are accumulated over $100$ independent simulations, as explained in the text.}
\label{fig2}
\end{figure}

To further investigate the consequences of group pressure, one calculated from the simulations the fraction of agents with extremist opinions ($o = \pm 1$) as a function of the group pressure for typical values of $\lambda$. We observed in Fig. \ref{fig1} that for the case $p=0$ we have $O=0$ for $\lambda<0.5$ (see also Eq. \eqref{eq5} for $p=0$). Thus, one exhibits in Fig. \ref{fig2} (left pannel) results for convictions $\lambda\geq 0.5$. One can see that, for sufficiently large values of $p$ the fraction of extremists goes to zero, even for the cases where agents present strong convictions like $\lambda=0.9$ or $1.0$. It is important to remember that the initial distribution of opinions is random ($[-1,1]$). Thus, for large $p$ a given agent $i$ has a tendency to align his/her opinion with the group's overall opinion. Considering all agents in their interactions, the consequence is that we have a wider distribution of opinions in the population. This result is also observed in the histograms of opinions (in the stationary states) exhibited in Fig. \ref{fig2} (right pannel), for $\lambda=1.0$ and distinct values of $p$.

\begin{figure}[t]
\begin{center}
\vspace{6mm}
\includegraphics[width=0.6\textwidth,angle=0]{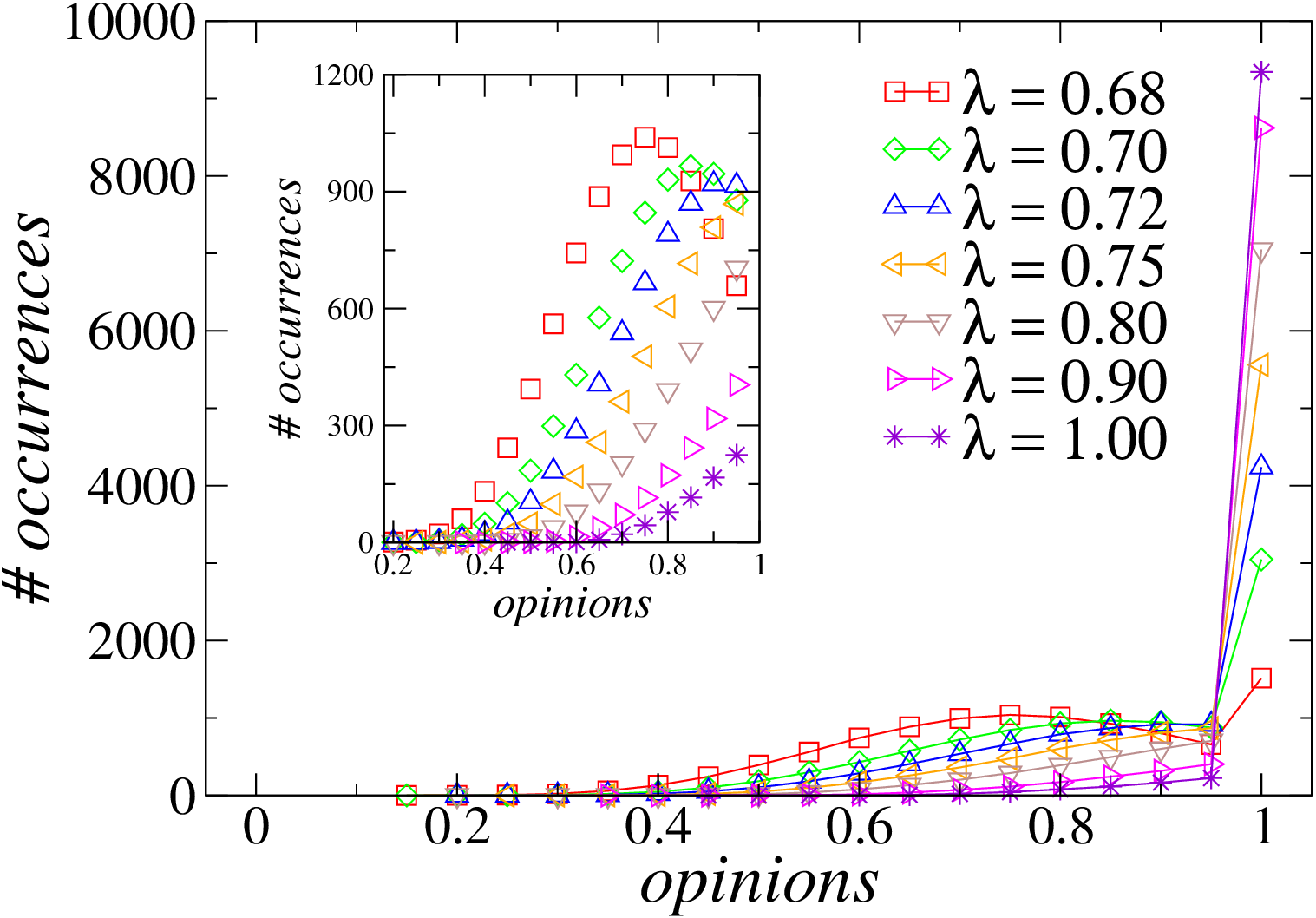}
\end{center}
\caption{(Color online) Histograms of opinions, in the stationary states, for $p=0.4$ and typical values of $\lambda>\lambda_{c}=2/3\approx 0.67$. In the inset we show a zoom of the main frame, excluding the extremist agents with majority opinions. The population size is $N=10^{4}$, and data are accumulated over $100$ independent simulations.}
\label{fig3}
\end{figure}

Let us elaborate about the construction of the histograms. Each histogram is obtained from $100$ independent simulations, for population size $N=10^{4}$. When there is unbalance of positive and negative opinions, we arbitrarily selected simulations with dominance of positive opinions to build the histograms \cite{jstat}. In this way, each histogram is representative of each single realization but with improved statistics. Instead, if we would have chosen the simulations with predominantly negative outcomes, the distribution would be the mirrored image of those shown in Fig. \ref{fig2}. Thus, in Fig. \ref{fig3} we exhibit another histogram, fixing $p=0.4$ and for increasing values of $\lambda$. In this case, we observe the emergence of extremist opinions for increasing values of agents' convicton, even for a intermediate value of $p$. Moderate opinions are observed for the smaller values of $\lambda$.

\section{Final remarks}   

\qquad In this work, we have studied a continuous opinion model based on kinetic exchange opinion dynamics. We also considered the presence of group pressure. The model take into account two rules, namely the 3-group interaction, that considers the agents' convictions $\lambda$, and the group pressure, quantified by a probability $p$. Stochastic parameters also drive the dynamics of interactios.

Our results suggest the occurrence of phase transitions at distinct critical points $\lambda_{c}(p)$. These critical points separates an absorbing phase, for $\lambda \leq \lambda_{c}$, where all agents become neutral (opinion $o=0$), from a ferromagnetic phase for $\lambda > \lambda_{c}$, where one of the opinion sides (positive or negative) disappears of the population in the steady states. From an analytical approach, we derived that $\lambda_{c}(p) = (2-p)/[4(1-p)]$. From numerical simulations we analyzed the distribution of opinions in the long-time limit, and our results suggest that the increase of group pressure leads to the decrease of extremism in the population. 

For future study, the properties of this model in various lattices and networks would be interesting, as well as the consideration of special agents like contrarians \cite{galam_cont}.

\section*{Acknowledgments}

The author acknowledges financial support from the Brazilian scientific funding agencies Conselho Nacional de Desenvolvimento Cient\'ifico e Tecnol\'ogico (CNPq) and Funda\c{c}\~ao de Amparo \`a Pesquisa do Estado do Rio de Janeiro (FAPERJ).

\end{document}